\documentstyle[graphicx,multicol,prl,aps,epsf]{revtex}
\begin{document}
\title{ Kondo tunneling through real and artificial molecules } 
\author{Konstantin Kikoin and Yshai Avishai}
\address{Department of Physics, Ben-Gurion University, Beer-Sheva 84 105,
Israel} \date{\today} \maketitle
\begin{abstract}
    When a cerocene molecule is chemisorbed on metallic substrate, or 
    when an asymmetric double dot is hybridized with itinerant electrons,  
    its singlet ground state crosses its lowly excited triplet state, 
    leading to a competition between the 
    Zhang-Rice mechanism of singlet-triplet splitting in a confined
    cluster and  the Kondo effect (which accompanies the
    tunneling through quantum dot under a Coulomb blockade restriction).  
   The rich physics of an underscreened $S=1$ Kondo impurity
    in the presence of low-lying 
    triplet/singlet excitations is exposed. Estimates of the 
    magnetic susceptibility and the electric conductance are presented. 
 \end{abstract}
\begin{multicols} {2}
     \noindent
\narrowtext
1. The problem of tunneling through a sequence of resonance levels
was formulated about three decades ago \cite{Car71}. 
At that stage, little attention was paid to the 
specific structure of the tunnel barrier. Nowadays, novel
experimental techniques enable the 
fabrication of artificial objects which 
carry some of the salient features of complex 
quantum systems existing in Nature, and to include them
as resonance barriers in  electron tunneling devices.
They manifest numerous unusual properties 
and might be regarded as  
important ingredients in future
microelectronics \cite{SCT92}.  Examples  are 
double quantum dot 
structures \cite{Taro95} \cite{Gup96}, atomic and 
molecular wires and bridges\cite{Joa97,Yan98}. 

At the same time,
new methods of tunneling microscopy make it possible to 
elucidate the properties
of single atoms and molecules adsorbed on a surface.  
The combination "nanotip -- atom/molecule -- substrate" 
is then a quantum system with 
exceptional resonance features and potential 
applications \cite{Joa97}. 

In the present work we expose the physics of tunneling  
through real and artificial molecules in which there is 
presumably a singlet ground state with an 
even number of electrons, which are {\it spatially separated into two 
groups with different degree of 
localization}.  
Electrons in the first group 
are responsible for strong correlation effects (Coulomb blockade),
whereas those in the second group are coupled to a metallic
reservoir. 
Hybridization with itinerant electrons 
result in transformation of  the nonmagnetic 
(singlet) ground state into a magnetic one. 

Possible real molecules are lanthanocene 
molecules Ln(C$_8$H$_8$)$_2$ with the ions Ln=Ce, Yb in a
cage formed by $\pi$-bonded carbon atoms \cite{Liu98}.  
In these molecules the electrons in a strongly correlated $f$-shell 
are coupled with loosely bound $\pi$ electrons. 
In an analogy with Zhang-Rice (ZR) singlet in Cu-O planes of high-T$_c$ 
perovskites \cite{Zhang88}, the ground state of
this molecule is a spin singlet combination 
$^1A_{1g}(f\pi^3)$ of an f-electron 
and $\pi$-orbitals, and the energy of the 
first excited triplet state 
$^3E_{2g}$ is rather 
small ($\sim 0.5$ eV). In the ytterbocene 
(hole counterpart of cerocene) the ground state with one $f$-hole is
a triplet, and the gap for a singlet excitation is tiny, $\sim 0.1$ eV.
The fullerene-like molecules doped with Ce or Yb form another family with
apparently similar properties. 
In all these systems there is no direct overlap between the strongly
correlated $f$-electrons and the metallic reservoir. However, these
electrons can
influence the tunnel properties of the molecule via covalent bonding
with the outer $\pi$-electrons which, in turn, are coupled to the 
metallic reservoir. 

Artificial candidates are double-dot 
structures (say $D_{1}$  and $ D_{2}$) in  
tunneling contact with each other,
but only $D_1$ is coupled with the metallic leads. 
The respective gate voltages are such that
$V_{g1}< V_{g2}$. Coulomb blockade
then prevents double charging of $D_{2}$,
so it can play 
the same role as $4f$ atom in molecular complexes described above. 
The dot $D_1$ donates the loosely bound electrons which 
contribute to the tunnel current \cite{Rok00}. 

The pertinent physics to be exposed below
is that of a competition between the 
ZR mechanism of singlet-triplet splitting in a confined
cluster and  
Kondo effect which accompanies the
tunneling through quantum dot under Coulomb blockade 
confinement \cite
{Glazr88b}. Usually, tunneling through quantum dot 
containing an even number of
electrons does not display a Kondo
resonance due to its
spin singlet ground state. Analysis of
conditions under which the singlet ground state changes
into a partially screened spin-one
Kondo state due to hybridization with metallic leads is
one of the goals of this study.

\noindent
2. A simple model which describes this type of molecules was
considered in \cite{Fulde}, 
hereafter referred to as a "Fulde molecule" (FM). 
It contains two electrons occupying a
potential well which is formed by deep and shallow valleys. 
The Hamiltonian of an isolated FM is 
\begin{equation}
H_{d}=\sum_{i}\sum_{\sigma }E_{i}n_{i\sigma }+V\sum_{i\neq j}d_{i\sigma
}^{\dagger }d_{j\sigma }+H_{corr}.
\label{1.2}
\end{equation}
Here $d_{i\sigma
}^{\dagger } $ creates a dot electron with spin $\sigma$ 
at valley $i=f,l$ 
and spin $\sigma$, while the
coupling constant $V=\langle d_{l}|V|d_{f}\rangle $ is the inter-well
tunneling integral.  There are two electrons in a neutral ground
state, and  
$H_{corr}=Qn_f(n_f-1)/2$
is the interaction term responsible for the
Coulomb blockade of charged states 
(here $n_f=\sum_\sigma d^\dagger_{f\sigma}d_{f\sigma}$). 
The energy difference  $\Delta= E_l - E_f$ is
postulated to exceed the overlap integral, $\beta=V/\Delta\ll 1.$
Two-electron states $|\Lambda\rangle$ of the FM are
classified as a ground state singlet $|S\rangle$, low-lying triplet exciton 
$|T0\rangle$, $|T\pm \rangle$ 
and high-energy singlet charge-transfer exciton 
$|L\rangle$. 
To order $\beta^{2}$ they are,
\begin{eqnarray}
|S\rangle & \approx & \alpha^2|s\rangle -\sqrt{2}\beta |ex\rangle, \nonumber \\
|T0\rangle & = & \frac{1}{\sqrt{2}}\sum_{\sigma}
d_{f\sigma }^{\dagger} d_{l-\sigma }^{\dagger}
|0\rangle, ~ |T\pm\rangle  =  d_{l\pm }^{\dagger }
d_{f\pm}^{\dagger}|0\rangle, \nonumber \\
|L\rangle & \approx & \alpha^{2}]|ex\rangle +\sqrt{2}\beta |s\rangle, 
\label{1.13}
\end{eqnarray}
where 
$|s\rangle  = \frac{1}{\sqrt{2}}\sum_{\sigma}\sigma d_{l-\sigma }^{\dagger }
d_{f\sigma }^{\dagger}|0\rangle, ~ 
ex\rangle = d_{l\uparrow }^{\dagger }d_{l\downarrow }^{\dagger }|0\rangle,$
and $\alpha^2=1-\beta ^{2}$.
\noindent
In this order, the energy levels $E_\Lambda$ 
are \cite{Fulde}: 
\begin{equation}
E_{S}  = \epsilon _{l}+\epsilon _{f}-2V\beta,  
E_{T}  =  \epsilon _{l}+\epsilon _{f}, 
E_{L}  =  2(\epsilon _{l}+V\beta ).  
\label{1.14} 
\end{equation}
The spin and charge branches
of excitation spectrum of FM are characterized by rather
different energy scales $E_T-E_S= \delta$ and $E_L-E_S\sim \Delta$, 
respectively. An interplay between 
Kondo triplet excitations (with some characteristic energy $\Delta_K$) and ZR 
triplet excitations is expected when $\delta \sim \Delta_K$
in the regime of 
Kondo resonance induced by tunneling to metallic reservoir 
\cite{foot1}. 

The tunneling problem is encoded in the Anderson Hamiltonian
which incorporates $H_{d}$, together with the
band Hamiltonian 
$H_{b}=\sum_{k\sigma}\epsilon_{k}c^{\dagger}_{k\sigma}c_{k\sigma}$ 
for the electrons in the leads, 
and the tunneling term  
$H_{t}=\sum_{ik\sigma}W_{i}c^{\dagger}_{k\sigma}d_{i\sigma}$. Here 
$c_{k\sigma}$ are operators for lead 
electrons and $W_{i=l,f}$ are tunneling matrix elements (assumed 
to be $k$ independent but strongly dependent on the dot valley 
quantum number $i$. It is henceforth assumed that $W_{f}=0$). It is 
convenient to express the dot operators $d_{i\sigma}$
in terms of Hubbard operators, 
$
X^{\Lambda \lambda}=|\Lambda \rangle \langle \lambda|
$.
Here $\Lambda =S,T,L$ stands for the neutral two-electron states 
(\ref{1.13}), 
and the index $\lambda =1\sigma ,3\sigma$ is reserved for the 
{\it charged} one and three electron states: 
$
|1\sigma \rangle \approx \alpha |f\sigma\rangle+\beta|l\sigma\rangle,~  
|3\sigma \rangle  \approx  d_{f\sigma }^{\dagger }|ex\rangle -\frac{V}
{Q-\Delta }d_{f\sigma }^{\dagger }d_{f\bar{\sigma}}^{\dagger}
d_{l\sigma }^{\dagger} |0\rangle.
$
The tunnel matrix elements  in the Hubbard representation are given as 
$W_{\sigma}^{\Lambda \lambda}=\langle 
k\sigma,\lambda|\hat W|\Lambda\rangle$, where $\hat W$ is 
the operator responsible for tunneling. The Anderson Hamiltonian then 
reads:
\begin{eqnarray}
H=\sum_{\Lambda }E_{\Lambda }X^{\Lambda \Lambda }+
\sum_{k\sigma }\epsilon _{k}c_{k\sigma }^{\dagger }c_{k\sigma }+\nonumber\\
\sum_{\Lambda \lambda }\left( W_{\sigma }^{\Lambda \lambda }
c_{k\sigma}^{\dagger }X^{\lambda \Lambda }+
\bar{W}_{\sigma }^{\Lambda \lambda}X^{\Lambda \lambda }c_{k\sigma }\right).   
\label{2.1}
\end{eqnarray}

\begin{figure}[htb]
\centering
\includegraphics[
height=0.4\textheight,angle=90,
keepaspectratio]{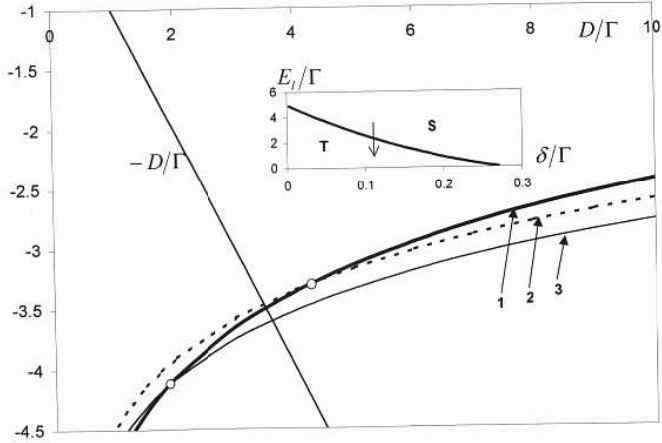}
\caption{
Scaling trajectories (Eq. \ref{3.5}) 
demonstrating cross-overs from
singlet to triplet ground state:
$\tilde{E}_T(D)$ (curve 1), $\tilde{E}_S(D)$ for 
$\delta/\Gamma=0.03, 0.1$ (curves 2,3, resp.) at 
$\Gamma_S/\Gamma_T=0.8,$~$D_0/\Gamma=10$. 
Inset: $S-T$ transition as a function of the level depth
$\varepsilon_F-\epsilon_l$ at fixed $\Delta$
}
\label{fig1}
\end{figure}

Using the Wigner-Eckart theorem, one can write 
$
W_{\sigma}^{\Lambda \lambda}=C_{\sigma\lambda}^{\Lambda}A_{\lambda},
$
where 
$C_{\sigma\lambda}^\Lambda$ are Clebsh-Gordan coefficients and 
$A_{\lambda}$ is the reduced matrix element.
In a given vector-coupling scheme the tunneling results in the following
transitions: $|S\rangle,|T0\rangle\leftrightarrow |1\sigma,
p\bar{\sigma}\rangle$; 
$|S\rangle,|T0\rangle\leftrightarrow |3\sigma, k\bar{\sigma}\rangle$; 
$|T\pm\rangle\leftrightarrow |1\pm, p{\pm}\rangle$; 
$|T\pm\rangle\leftrightarrow| 3\pm, k\mp\rangle$. 
Here $p\sigma$ and $k\sigma$ are, respectively, the states with 
an excess electron (and hole) above (below)
the Fermi level of the lead. 
Let us focus on the case where the
Coulomb blockade eliminates the three electron states 
$|3\sigma\rangle$ and consider the tunnel coupling involving only the
states $|1\sigma\rangle$.
The non-zero tunnel matrix elements are 
\begin{equation}
W_{\pm}^{T\pm}=W,~~ 
W_{\mp}^{T0}=\frac{1}{\sqrt{2}}W,~~
W_{\mp}^{S}=\pm\frac{\alpha^2}{\sqrt{2}} W , 
\label{2.6}
\end{equation}
where $W=\alpha W_{l}$.
The energy
costs of these transitions are 
\begin{eqnarray}
E_{1p,S} & = & \epsilon_p-\epsilon_l+\beta V,~~
E_{1p,T}=\epsilon_p-\epsilon_l, 
\label{2.5a}\\
E_{3k,S} & = & \epsilon_l+4\beta V+\widetilde{Q}-\epsilon_k,~~
E_{3k,T}=\epsilon_l+2\beta V +\widetilde{Q}-
\epsilon_k, 
\nonumber
\end{eqnarray}
where $\widetilde{Q}\approx Q[V^2/(Q-\Delta)^2].$ 
\noindent
3. We study the interplay between the singlet and triplet levels 
of the double quantum dot by the renormalization group method
following the general line of "poor man's scaling" approach to the  
Anderson model \cite{Hald78}. 
The renormalized levels $\tilde{E}_{\Lambda}$ 
are determined by the equations
\begin{equation}
d\tilde E_\Lambda/d\ln D=\Gamma_\Lambda/\pi.
\label{3.5}
\end{equation}
Here $\Gamma_{\Lambda}$ 
are the tunnel coupling constants,
\begin{equation}
\Gamma_{T}=\Gamma \equiv \pi \rho_0 W,~~~
\Gamma_S=\alpha^2 \Gamma_T,~~~
\rho_0\sim D^{-1}.
\label{3.3}
\end{equation}  
Integrating $(\ref{3.5})$ under the conditions 
$\tilde{E}_\Lambda(D_0)=E_\Lambda,~
\Gamma_\Lambda=const$, 
we find the scaling invariants $E_\Lambda^*$ which 
determine the scaling trajectories
\begin{equation}
E_\Lambda^*=E_\Lambda - \frac{\Gamma_\Lambda}{\pi}
\ln\left(\frac{\pi D}{\Gamma_\Lambda}\right)
\label{3.6}
\end{equation}
The level $\epsilon_f$ is taken to be close to the bottom of 
the conduction band \cite{foot1}, 
so that scaling does not significantly affect it.
It is then
subtracted from the energies $E_T$ and $E_S$.
Now we see that the energies $E_\Lambda$ 
decrease together with $D$.
Since $\Gamma_T>\Gamma_S$, the phase trajectory $E_T(D,\Gamma_T)$ 
should cross 
that of $E_S(D,\Gamma_S)$ at 
a certain point. Thus, quite remarkably, 
there is a crossover from 
singlet to triplet ground state of 
the FM due to tunnel contact with 
metallic leads. The crossing 
point can be 
estimated from eqs. (\ref{3.6}):
$\pi\delta^*\approx(\Gamma_T-\Gamma_S)\ln(\pi D/\Gamma),
$
or, referring to the bare parameters, the value $\widetilde{D}$ of 
renormalized bandwidth corresponding to this crossing point is 
$
\widetilde{D}=D\exp\left(-\pi\Delta/\Gamma\right).
$
Another important crossing point is the energy $\bar{D}=a \bar{\epsilon}_l$ 
($a \gtrsim 1$)
where the scaling of $\epsilon_l$ also stops, the charge fluctuations become 
irrelevant, and one reaches the 
Schrieffer-Wolff limit where only spin 
fluctuations are responsible for scaling of the model Hamiltonian 
\cite{Hald78}.
This energy is determined by the 
equation
\begin{equation}
\bar{D}=(\Gamma/\pi)\exp
\left(\pi(|\bar{\epsilon_l}-\epsilon_l|)/\Gamma\right).
\label{3.7b}
\end{equation}
If $\widetilde{D}>\bar{D}$, 
the Schrieffer-Wolff regime is reached
{\it after} crossover from singlet to triplet
 ground state of the FM, and
a Kondo type resonant tunneling is feasible. 
In the opposite case there is a singlet ground state and  
a soft triplet exciton (see figure \ref{fig1}). 
And yet, 
the $S=1/2$ Kondo regime is still accessible once a  
properly tuned external magnetic field 
is applied \cite{APK}. 
The novel feature here is that $S\to T$ crossover 
can be induced by an upward shift of the 
dot level $\epsilon_l$ relative to $\varepsilon_F$ by a 
suitable gate voltage 
(Fig. 1, inset).

We focus on
the physically richer case of triplet solution $\bar{\delta}<0$. 
For a two-electron FM
the Schrieffer-Wolff transformation \cite{Hald78} 
projects out the states $|\lambda,k\sigma\rangle$ and maps 
the Hamiltonian $H$
onto an effective Hamiltonian $\widetilde{H}$ acting in a subspace of 
two-electron configurations $|\Lambda\rangle$ and reduced conduction band, 
$
\widetilde{H}=\widetilde{H}^S+\widetilde{H}^T+\widetilde{H}^{ST},
$
\begin{eqnarray}
\widetilde{H}^S & = & \widetilde{E}_S X^{SS}
+J^S\sum_{\sigma}
X^{SS}c^\dagger_{\sigma}c_{\sigma}
\nonumber \\
\widetilde{H}^T & = & \widetilde{E}_T \sum_\mu X^{\mu\mu}+
J^T {\bf S\cdot s}  
+\frac{J_T}{2}\sum_{\mu\sigma}X^{\mu\mu}c^\dagger_{\sigma}c_{\sigma},
\nonumber \\
\widetilde{H}^{ST} & = & J^{ST}\left({\bf P}\cdot{\bf s}\right) .
\label{3.14}
\end{eqnarray}
$(\mu=T0,T\pm)$.
The local electron operators are defined as usual
$
c_\sigma=\sum_k c_{k\sigma};~  
{\bf s}=2^{-1/2}\sum_{kk'}\sum_{\sigma\sigma'}
c^\dagger_{k\sigma}\hat{\tau}c_{k'\sigma'};$
$\hat{\tau}$ are the Pauli matrices. 
The singlet and triplet states are now intermixed, and the spin properties 
of FM are characterized by the vector operators ${\bf S}$ and 
${\bf P}$ in accordance with the dynamical symmetry of spin rotator: 
\begin{eqnarray}
S^+ & = & \sqrt{2}\left(X^{+ 0}+X^{0-}\right), 
S^-  =  \sqrt{2}\left(X^{0+}+X^{- 0}\right),\nonumber \\
S^z & = & X^{++}-X^{--},~
P_z  =  -\left(X^{0S}+X^{S0}\right),
\label{3.9c}\\
P^+ & = & \sqrt{2}\left(X^{+ S}-X^{S-}\right),\;
P^-  = \sqrt{2}\left(X^{S+}-X^{- S}\right),\nonumber 
\end{eqnarray}
These operators obey the moment algebra ($i=x,y,z):$
\begin{equation}
[P^i,P^j] = i\varepsilon_{ijk}S^k,~ 
[P^i,S^j] = i\varepsilon_{ijk}P^k,~
{\bf S\cdot P} = 0. 
\label{3.11}
\end{equation}
and the Casimir operator is $S^2+ P^2 =3.$ Surprisingly, 
this special representation of $O(4)$ played an important 
role in particle physics many years ago \cite{DGN}.
The effective exchange integrals are  
\begin{eqnarray}
J^T=-\frac{2|W_l|^2}{\Delta_T},~
J^S & = & -\frac{\alpha^2|W_l|^2}{\Delta_S},~
J^{ST}=-\frac{\alpha|W_l|^2}{\sqrt{2}\bar{\Delta}},\nonumber \\
\bar{\Delta}^{-1} & = & \Delta_T^{-1}+\Delta_S^{-1},
\label{3.10}
\end{eqnarray}
in which $\epsilon_{k}$ is replaced by $\epsilon_{F}$ in the 
denominators, that is, 
$\Delta_\Lambda=\epsilon_F-\epsilon_\Lambda(\bar{D})$, and 
$\epsilon_\Lambda(\bar{D})$ are the positions of the scaled level  
$\epsilon_l$ on the flow diagram of Fig. \ref{fig1}. 
Thus, the pertinent physics is that of an underscreened Kondo impurity 
\cite{Noz80} in the presence of potential scattering and low-lying 
triplet/singlet excitations. 
A similar model was considered recently in Ref.\cite{Eto00} 
studying the physics of
tunneling through a vertical quantum dot in magnetic field 
\cite{Sas00}. In that case, the electron orbital motion in 
a plane perpendicular to the axis of the dot is characterized by the same
quantum number both in the dot and in the leads \cite{Pust00}, and 
two orbitals participate in the $S-T$ transitions. The problem can 
then be mapped onto a special version of the two-impurity Kondo model. 

Following \cite{Hald78} we now apply the
"poor man scaling approach" \cite{Anders70} to the Hamiltonian $\tilde{H}$
(\ref{3.14}). Neglecting the irrelevant potential scattering 
phase shift \cite{Noz78} and using
the above mentioned cutoff procedure, 
a system of scaling equations is obtained,(cf. \cite{Eto00})
\begin{equation}
dj_1/d\ln d = 
-\left[(j_1)^2+(j_2)^2\right],~
dj_2/d\ln d~  =  -2j_1j_2
\label{3.13}
\end{equation}
(here $j_1=\rho_0J^T, j_2=\rho_0J^{ST}, d=\rho_0D$).
The corresponding RG flow diagram has the fixed point $j_1=\infty $,
but the resulting Kondo temperature $T_K(\bar\delta)$ turns out 
to be a sharp function of $\bar\delta$ 
\cite{Eto00}. It is maximal  
when the $T,S$ states are quasi degenerate, 
$\bar\delta \ll T_K(\bar\delta)$. The scaling in this case is governed 
by the effective
integral $j_+=j_1+j_2$, and the system (\ref{3.13}) 
is reduced to a single equation
\begin{equation}
dj_+/d\ln d  =  -(j_+)^2
\label{3.15}
\end{equation}
with $T_{K0}=\bar{D}\exp(-1/j_+)$.
In the opposite limit
$\bar{\delta} \gg T_K(\bar{\delta})$ the scaling of $J^{ST}$ stops 
at $D\simeq \bar{\delta}$. 
 Then
$j_{1,2}(\bar\delta)=j_{1,2} \ln^{-1}
\left(\frac{\bar\delta} {T_{k0}}\right)$ and 
$T_{K}(\bar{\delta})=\bar\delta \exp\left[-1/j_{1}(\bar\delta)\right] 
\ll T_{K0}$. 
The singlet ground state $S$ with
zero $T_K$ is realized 
when $\bar{\delta}<0, |\bar{\delta}|>T_{K}(\bar{\delta})$. 

\noindent 
4. The salient features of FM stem from the qualitative dependence 
of its ground state and low-energy spectrum 
on the coupling constants $V$ and $W_l$. 
The unusual singlet-triplet crossing should show up 
in the magnetic properties
of adsorbed molecules and tunnel transparency of asymmetric double
quantum dots. 

According to quantum chemical calculations of the energy spectrum of 
isolated cerocene molecule, the Van Vleck paramagnetic contribution of
$S-T$ excitations is too weak to overcome the 
Larmor diamagnetic contribution 
of C$_8$H$_8$ rings \cite{Liu98,Fulde}. 
This situation can drastically change
for a FM  adsorbed on a metallic layer. 
The fixed point $j_1=\infty $ corresponds to the scattering phases 
$\eta_\sigma(\epsilon_F)=\pi/2.$ 
In the case of adsorbed FM this means that 
the  molecule has a residual spin 1/2 which interacts ferromagnetically
with the conduction electrons \cite{Noz80}.  
The temperature dependence of magnetic susceptibility $\chi(T)$ 
is predetermined by 
the energy  parameters $\bar{\delta}$ and $T_K(\bar{\delta})$. In particular, 
$\chi(T)$ conserves its Curie-like character down to the lowest temperatures
when $\bar\delta<0$, 
$|\bar{\delta}|\gg T_K$ 
Then at $T\ll T_K$ the underscreened FM remains
paramagnetic,  and its susceptibility is 
\begin{equation}
\chi(T)=\chi_0(T)[1-Z(T/T_K)]
\label{4.10}
\end{equation}
Here
$\chi_0=3C/4T$, $C=(g\mu_b)^2$, and 
$Z(x)$ is the invariant coupling
function (solution of the Gell-Mann -- Low equation, see \cite{Faw81}). 
The triplet spin state is restored at $T > T_K $ .  
In this regime the Kondo corrections as well as admixture of 
singlet state can be calculated
by perturbation theory, with the result,
\begin{eqnarray}
\chi(T) & = & 
\frac{2C[3-\exp(-\bar{\delta}/T)]}{3T}
\left(
1-1/\ln
\frac{T}{T_{K}(\delta)}-j_2\ln\frac{\bar{D}}{\bar{\delta}}
\right), \nonumber \\
\chi(T) & = &
\frac{2C}{T[3+\exp(-\bar{\delta}/T)]}
\left(1-1/\ln\frac{T}{T_{K0}}\right),
\label{4.9}
\end{eqnarray}
respectively for $\bar{\delta} \gg T_{K0}$ and 
$\bar{\delta} \ll T_{K0}$.
In the case of artificial FM the resonance scattering phase means perfect
tunneling transparency of the quantum dot at $T=0$ and a logarithmic fall 
off at high temperatures.
To calculate the 
tunneling transparency of FM sandwiched between two leads, one should 
add an index $n=L,R$ to the operator $c_{nk\sigma}$ and 
switch to the standing wave basis \cite{Glazr88b} 
$
\sqrt{2}c_{k\sigma \pm} =c_{Lk\sigma}\pm c_{Rk\sigma}
$
(in a symmetric configuration $W_{iL}=W_{iR}$).
Then only the wave (+) is involved in tunneling, and 
the zero bias anomaly in the differential conductance
$G(T)$ (due to Kondo cotunneling) in the weak
coupling regime $T>T_K$ is found  as in (\ref{4.9}):
\begin{eqnarray}
G/G_0 & = & 2\ln^{-2}[T/T_{K}(\delta)]+j_1j_2^2\ln(\bar{D}/\bar{\delta}),
\nonumber \\
G/G_0 & = & 3\ln^{-2} [T/T_{K0}]
\label{4.1}
\end{eqnarray} 
respectively for the two limiting cases $\bar{\delta}\gg T \gg T_{K\delta}$ and 
$T\gg T_{K0}\gg \bar{\delta}$.
Here $G_0=4\pi e^2/\hbar$. Again the maximum effect is achieved in a 
nearly degenerate case. At $T\to 0$ the conductance tends to the
unitarity limit. 

In conclusion, the interplay between
ZR-type coupling in real and/or artificial molecules and
Kondo coupling between molecules and metallic reservoir
may result in a crossover from a singlet spin state in a
weak-coupling regime to an underscreened $S=1$ state at zero $T$.
The onset of Kondo regime in double
quantum dot with even occupation can be driven
either by a magnetic field or by a gate voltage.

\noindent{\bf Acknowledgmet} This research is supported by    
Israeli S.F. grants "Nonlinear Current Response of Multilevel Quantum        
Systems" and "Strongly Correlated Electron Systems in Restricted 
Geometries", DIP grant ``Quantum Electronics in Low Dimensions'' 
and BSF grant ``Dynamical Instabilities in Quantum Dots''. 
We are grateful to I. Krive for valuable discussions and to I. Kikoin
for assistance in numerical calculations.

\end{multicols}

\begin{thebibliography}{99}
\bibitem{foot1} To satisfy the expectation $\delta\sim T_K$ the gap $\Delta$ 
should be at least of order of $D$
\end{thebibliography}
\end{document}